\documentclass[conference]{IEEEtran}
\IEEEoverridecommandlockouts

\usepackage{cite}
\usepackage{amsmath,amssymb,amsfonts}
\usepackage{algorithmic}
\usepackage{graphicx}
\usepackage{booktabs}
\usepackage{textcomp}
\usepackage{colortbl}
\usepackage[table]{xcolor}
\def\BibTeX{{\rm B\kern-.05em{\sc i\kern-.025em b}\kern-.08em
    T\kern-.1667em\lower.7ex\hbox{E}\kern-.125emX}}

\begin{document}

\title{
Optimizing Compilation for Distributed Quantum Computing via Clustering and Annealing
}

\author{\IEEEauthorblockN{
Ruilin Zhou\IEEEauthorrefmark{2}\IEEEauthorrefmark{1}, 
Jinglei Cheng\IEEEauthorrefmark{4}\IEEEauthorrefmark{1},
Yuhang Gan\IEEEauthorrefmark{2}, 
Junyu Liu\IEEEauthorrefmark{4},
Chen Qian\IEEEauthorrefmark{2},}
\IEEEauthorblockA{\IEEEauthorrefmark{4}University of Pittsburgh \
\IEEEauthorrefmark{2}University of California, Santa Cruz\\
\IEEEauthorrefmark{1}Authors contributed equally to this research. 
}
}

\maketitle

\begin{abstract}
Efficiently mapping quantum programs onto Distributed quantum computing (DQC) are challenging, particularly when considering the heterogeneous quantum processing units (QPUs) with different structures.
In this paper, we present a comprehensive compilation framework that addresses these challenges with three key insights: exploiting structural patterns within quantum circuits, using clustering for initial qubit placement, and adjusting qubit mapping with annealing algorithms.
Experimental results demonstrate the effectiveness of our methods and the capability to handle complex heterogeneous distributed quantum systems. Our evaluation shows that our method reduces the objective value at most 88.40\% compared to the baseline. 
\end{abstract}


\section{Introduction}
\label{sec-intro}
The development of quantum hardware has been rapid and groundbreaking across different quantum platforms, including trapped ion~\cite{Bruzewicz2019}, superconducting~\cite{clarke2008superconducting}, neutral atom~\cite{henriet2020quantum}, quantum dot~\cite{loss1998quantum}, and photonics~\cite{slussarenko2019photonic}. 
On the software side, significant progress has been made in developing quantum algorithms~\cite{montanaro2016quantum,nielsen2010quantum}, quantum error correction~\cite{lidar2013quantum}, and compilation tools~\cite{ge2024quantum}. 
These advancements have collectively brought us closer to realizing practical quantum advantages in various tasks, including quantum chemistry~\cite{levine2014quantum}, quantum search~\cite{grover1996fast}, quantum machine learning~\cite{biamonte2017quantum,schuld2015introduction}, and combinatorial optimization~\cite{farhi2014quantum}.

As we transition from near-term intermediate-scale quantum (NISQ) computing to fault-tolerant quantum computing (FTQC), the size of quantum computers has reached over one thousand qubits~\cite{ibm2023condor}. 
Scaling quantum computers further for practical tasks requires DQC with multiple Quantum Processing Units (QPUs). 
However, this approach introduces several challenges. 
One of the most critical challenges is the mapping problem, which is assigning logical qubits to physical qubits in a way that accounts for the connectivity constraints and communication overhead between QPUs.
Traditional mapping techniques~\cite{li2019tackling,zulehner2018efficient,zhang2022qubit} are designed for single QPU and not optimized to address the overhead introduced by inter-QPU communication, which are implemented as remote gates and quantum teleportation. 
Therefore, there is a pressing need for new mapping strategies that are specially designed for distributed quantum systems.
Additionally, the mapping algorithms for distributed quantum system must be scalable, as they target larger and deeper quantum programs with more qubits than those in the NISQ era.

Recent works~\cite{wu2022autocomm,mao2023qubit,wu2023qucomm,ferrari_compiler_2021-1,caleffi_distributed_2022} have proposed various methods to optimize the mapping and compilation process for DQC. 
For instance,~\cite{wu2022autocomm} utilized burst communication to reduce the impact of remote gates on overall program latency, and~\cite{wu2023qucomm} explored qubit re-allocation to minimize qubit movement across different QPUs. 
However, existing compilation methods often reply on simplified assumptions: they assume identical configurations across all QPUs, including uniform qubit counts, identical error rates, consistent internal connectivity, and homogeneous communication resources. 
Such assumptions of homogeneity cannot accurately represent the reality of distributed quantum systems, where heterogeneity is inherent and often unavoidable.
Different QPUs within a cluster may vary significantly in their number of computational and communication qubits, and possess different hardware topology. 
This kind of heterogeneity is particularly unavoidable when integrating QPUs from different evolving generations, which can be a common scenario in early-stage distributed quantum computing systems.

\begin{figure}
\vspace{-3mm}
    \centering
    \includegraphics[width=0.9\linewidth]{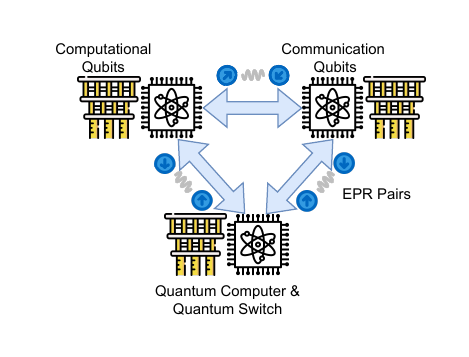}
    \caption{Overview of our DQC framework. The system architecture consists of heterogeneous QPUs with varying qubit counts and topologies, interconnected through quantum switches that manages EPR pair distribution for remote operations. The quantum switch enables efficient entanglement resource management and teleportation between QPUs.}
    \label{fig:teaser}
\vspace{-3mm}
    
\end{figure}

At the hardware level, clusters are composed of QPUs with different specifications and capabilities. 
At the program level, quantum circuits also have non-uniform structures and different gate densities across different regions\cite{tan2023quct,lourens2023hierarchical}.
The methods of previous works do not consider the gate patterns within quantum programs.
By gate patterns, we mean that gates tend to concentrate on certain qubits rather than being evenly distributed across the entire program. 
These factors indicate that previous DQC compilation and mapping methods, which focus solely on minimizing communication overhead, may not be suitable for future DQC hardware. 
Instead, we need compilation and mapping methods that can systematically consider these factors—heterogeneous QPUs with different qubit numbers and topologies, the evolving landscape of quantum interconnects, and the structural patterns within quantum circuits—to enhance computational efficiency and scalability in distributed quantum systems.

In this work, we address previous challenges of efficient quantum circuit mapping in distributed quantum computing. The method comprises three key steps: partitioning quantum circuits into segments based on gate patterns, performing initial qubit allocation through time-aware clustering, and refining this allocation using simulated annealing. The annealing-based optimization effectively balances the trade-off between local and remote operation overheads while accounting for the heterogeneity of QPUs with different qubit numbers and topologies. By systematically incorporating circuit patterns, quantum interconnect capabilities through our switch design, and cluster heterogeneity, our framework achieves efficient mapping of logical qubits to physical qubits in distributed quantum systems.

Our contributions are threefold:
\begin{itemize}
    \item We introduce a pattern-aware framework for distributed quantum circuit mapping that effectively handles heterogeneous QPU clusters with varying qubit counts and topologies.
    \item We develop a quantum switch-based architecture combined with simulated annealing optimization that efficiently manages entanglement resources and balances the inherent trade-off between local computation and remote communication overheads.
    \item We demonstrate significant improvements(88.40\%) compared to baselines in distributed quantum circuit execution through our comprehensive approach, achieving better resource utilization and reduced overall overhead compared to existing methods.
\end{itemize}

The remainder of this paper is organized as follows: Section~\ref{background} provides the necessary background on DQC. Section~\ref{motivation} presents our motivation and discusses the challenges in mapping quantum circuits to heterogeneous QPU clusters. Section~\ref{method} details our method, including the quantum switch architecture, pattern-aware circuit partitioning, and simulated annealing-based optimization. Section~\ref{evaluation} evaluates our approach through comprehensive experiments and comparisons. Finally, Section~\ref{conclusion} concludes the paper.

\begin{figure*}
    \centering
    \includegraphics[width=0.9\linewidth]{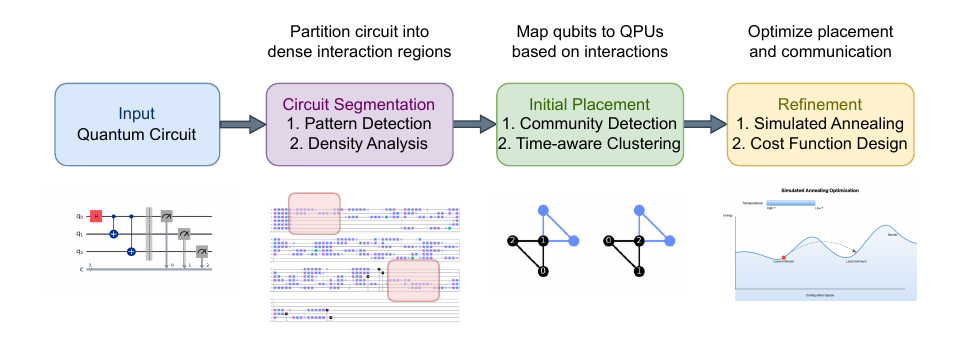}
    \caption{The workflow of our proposed method. There are three major steps: circuit segmentation, initial placement and refinement. The circuit segmentation is to identify circuit patterns that are regions of dense quantum gates. The initial placement is using community detection to minimize the possible remote gates between QPUs and provide a good starting point for further optimization. And the refinement is using simulated annealing to adjust the qubit allocation with a cost function that considers inter- and intra- QPU costs as in the near future the intra-QPU overhead can not be neglected compared with inter-QPU operations.}
    \label{fig:enter-label}
\end{figure*}

\section{Background}
\label{background}

\subsection{Basics of Quantum Computing}
The fundamental unit of information in quantum computing is the quantum bit (qubit). 
And state vectors are used to represent qubits. 
The states $|0\rangle = \left[ \begin{array}{c} 1 \\ 0 \end{array} \right]$ and $|1\rangle = \left[ \begin{array}{c} 0 \\ 1 \end{array} \right]$ correspond to the classical bits 0 and 1, respectively. 
A qubit can exist in a linear combination of these states, expressed as $|\psi\rangle = \alpha|0\rangle + \beta|1\rangle$, where $\alpha$ and $\beta$ are complex numbers such that $|\alpha|^2 + |\beta|^2 = 1$. 
The state of a qubit can be manipulated using single-qubit gates such as Hadamard gate $H$ that creates an equal superposition state from a basis state, or Pauli gates $X, Y\ $ and $\ Z$, etc. 
Upon measurement, the qubit collapses to the state $|0\rangle$ with probability $|\alpha|^2$, or to the state $|1\rangle$ with probability $|\beta|^2$. 
Two or more qubits can be entangled using two-qubit Control-NOT (CNOT) gates. 
Additionally, quantum gates that operate on three or more qubits can be decomposed into combinations of single- and two-qubit gates \cite{barenco1995elementary}. 
Most existing quantum programs and quantum algorithms use the circuit model, a sequence of gates acting on different qubits, to represent.

\subsection{Distributed Quantum Computing}
\label{subsec-dqc}
Recent advances in quantum networks and quantum interconnects have enabled distributed quantum computing, where a large quantum program can be split and distributed into different quantum computing nodes\cite{pompili2021realization, hermans2022qubit, magnard2020microwave, li2024high}. 
Similar to classical distributed computing, distributed quantum computing relies on remote communications. 

If two qubits are in one of the Bell states, they form an EPR pair \cite{nielsen2010quantum}. 
In DQC, each qubit of one EPR pair can be located in different quantum nodes to create a remote EPR pair; and remote communication involves establishing a remote EPR pair between two nodes and then consuming this EPR pair as a communication resource to transfer quantum info. 
Most previous works on distributed quantum computing have utilized one of two methods to perform remote gates: 
\begin{itemize}
    \item Using a cat-entangler and cat-disentangler \cite{yimsiriwattana2004generalized}
    \item Using quantum teleportation \cite{nielsen2010quantum}
\end{itemize}
Both methods consume EPR pairs as communication resources. 
In the DQC configuration, each node contains two types of qubits: communication qubits and computational qubits. 
computational qubits store quantum states and execute quantum gates, while communication qubits generate remote EPR pairs and then perform remote gates.

\section{Motivation}
\label{motivation}

\subsection{heterogeneous QPU}
One of the common assumptions in previous DQC work is that DQC will be implemented with a homogeneous all-to-all connected topology of QPUs, ignoring the more general case: heterogeneous clusters of QPUs. 
In realistic setups, heterogeneity can exist in several critical aspects:
\textbf{Different QPU topologies}: As in today's IBM Quantum Cloud platform\cite{ibm2023roadmap}, which hosts QPUs with different topologies. 
This variation impacts how quantum circuits are mapped onto the hardware, as certain qubit interactions may not be directly available, requiring additional overhead to implement quantum operations. 
\textbf{Different numbers of computational and communication qubits}: Some QPUs may have more qubits dedicated to computation, while others allocate more qubits for communication. 
This disparity affects the distribution and execution of quantum algorithms across the network. And it requires adaptive resource allocation strategies. 
\textbf{Inconsistent qualities of quantum links affected by physical distance and environmental interference}: Quantum links between QPUs will drift in fidelity due to factors including attenuation over distance and environmental noise\cite{gyongyosi2022advances}. 
This leads to unstable communication reliability and performance, which must be addressed in algorithm design. 
\textbf{Diverse error rates originated from hardware implementations and error correction capabilities}: Different QPUs will exhibit diverse error rates due to differences in qubit decoherence time and their properties out of the factory\cite{google2023suppressing}. This impacts the overall accuracy of computations and requires specially designed error mitigation techniques.

\subsection{Faster Quantum Interconnects}
Recently, we have seen improvements in building faster and more reliable quantum interconnects from both hardware advancements and theoretical analysis. 
One example is shown in~\cite{li2024high}, where a Bell pair rate approaching $10^5~\mathrm{s}^{-1}$ with a fidelity around 0.9999 is reported. 
Another example is presented in~\cite{pattison2024fast}, where a method to generate high-fidelity Bell pairs with distillation is provided. 
From these advancements, we can anticipate that in the near future, the time required to generate high-quality and low-overhead Bell pairs will be comparable to that of local operations. 
Therefore, under this circumstances, compilation methods in DQC that focus solely on minimizing communication overhead or the number of EPR pairs will not be the optimal. 
Instead, \textbf{we need methods to systematically explore the trade-off between local operation overhead and remote operation overhead, especially when under heterogeneous assumptions.}
We incorporate a quantum switch in our system model. The switch acts as a centralized manager for entanglement resources, coordinating the generation and distribution of EPR pairs between different nodes in the network. This centralization simplifies the modeling and optimization of resource consumption, particularly for remote operations like teleportation and cat-entanglement operations, each of which consumes one EPR pair. The quantum switch abstraction also allows us to focus on higher-level optimization decisions while maintaining a clear accounting of the entanglement resources required for distributed execution.

\subsection{Patterns in Quantum Circuits}

In addition to the challenges posed by heterogeneous QPU architectures and advancements in quantum interconnects, the presence of patterns or unique structures within certain quantum circuits also influences the compilation of quantum circuits onto DQC. 
Many quantum algorithms—such as the Quantum Fourier Transform (QFT), quantum adders, and variational circuits used in the Variational Quantum Eigensolver (VQE)—have regular and repetitive structures. 
It is worth noting that even in large quantum circuits, there are instances where only a limited number of qubits are actively involved at a given time window. 
In the implementation of quantum algorithms, some quantum circuits are inserted with repeated substructures.

Recognizing and leveraging these patterns allow for more efficient mapping of quantum operations onto DQC, which minimizes resource consumption and reduces communication overhead. 
Exploring these circuit patterns enables the compiler to match specific sub-circuits with certain QPUs that have the most suitable topology and characteristics. 
Therefore, when designing compilation methods, it is crucial to capture this structural information to develop strategies that effectively balance the trade-offs between local and remote operation overheads under heterogeneous assumptions.

\section{Method}
\label{method}

In this section, we present our methods for compiling distributed quantum circuits across multiple QPUs.
Our methods have three key steps: \textbf{circuit segmentation}, \textbf{initial placement}, and \textbf{simulated annealing optimization}. 
During the circuit segmentation phase, we identify groups of highly interacting qubits within specific time windows and vertically partition the circuit to isolate these interactions. Building on the segmentation results, we encode temporal information and apply time-aware clustering to establish an initial qubit mapping, ensuring that qubits with frequent near-future interactions are positioned in close QPUs. Here, the distance of the QPUs is determined by the topology of how QPUs are connected. This assumption is different from previous works where QPUs are fullly connected. Finally, we use simulated annealing to refine the qubits allocation by minimizing both local and remote operational overheads while also considering the heterogeneous characteristics of different QPUs and quantum links. 

\begin{figure}
    \centering
    \includegraphics[width=\linewidth]{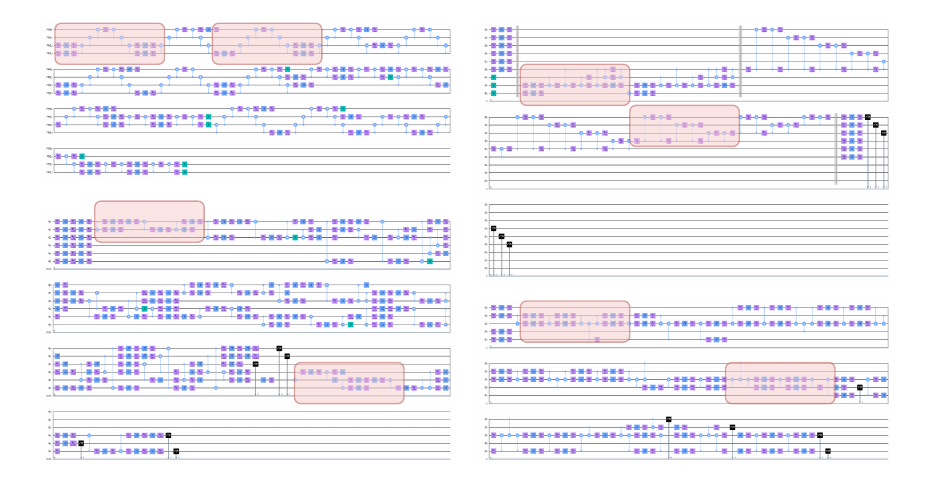}
    \caption{Visualization of gate distribution patterns in quantum circuits. The pink highlighted regions indicate areas of concentrated quantum operations, showing recurring gate patterns and non-uniform gate distributions across qubits. This natural clustering of quantum operations suggests opportunities for circuit optimization through pattern-aware compilation.}
    \label{fig:pattern}

\end{figure}
\subsection{Circuit Segment}

The architecture of quantum circuits frequently exhibits patterns and repeated structural motifs. These patterns have temporal segments where only specific qubit subsets actively participate in quantum operations or as recurring interaction patterns that appear throughout circuit execution. Such structural properties have significant implications for distributed quantum computing systems and are often ignored by previous work, particularly when distributing computations across heterogeneous QPUs. This structural pattern exploitation is demonstrated in Figure.~\ref{fig:pattern}, where we show different patterns of two types of circuits. We aim to segment the circuit into intervals where interactions among these qubits are denser. To achieve this, we compute an interaction density metric over time for each qubit to quantify its activity level in two-qubit gates. 
At each layer $l$ of the circuit, we calculate the cumulative number of two-qubit gates involving qubit $q$ up to that layer, denoted as $D_q(l)$. 
The interaction density $\rho_q(l)$ for qubit $q$ at layer $l$ is then defined as $\rho_q(l) = \frac{D_q(l)}{l}$. 
To capture the interaction periods and mitigate possible transient fluctuations, we apply a sliding window of size $w$ to the interaction densities, computing the windowed average interaction density  
\begin{equation}
    \bar{\rho}q(l) = \frac{1}{2w} \sum{i = l - w}^{l + w} \rho_q(i) 
\end{equation}
At each layer $l$, we identify the set $S_l$ consisting of the top $k$ qubits with the highest windowed densities  $\bar{\rho}q(l)$. 
To detect transitions in interaction patterns, we measure the similarity between consecutive sets $S_l$ and $S{l-1}$  with the Jaccard index $J(S_l, S_{l-1}) = \frac{|S_l \cap S_{l-1}|}{|S_l \cup S_{l-1}|}$, and initiate a new segment at layer $l$ when  $J(S_l, S_{l-1}) $ falls below a predefined threshold $\theta$ where specific $w, k, and \theta$ depends on the width and depth of the circuit. 

\subsection{Initial Placement}
Given the circuit segmentation in the previous step, we determine the initial placement of qubits onto QPUs with time-aware clustering. 
The key idea is to construct an interaction graph that reflects the frequency of qubit interactions and incorporates temporal information.
Temporal information is implemented by assigning greater importance to near-future interactions. 
We begin by constructing a global interaction graph $G = (V, E)$ that has all circuit segments. 
Each vertex $v_i \in V$ represents a logical qubit $q_i$, and each edge $e_ij \in E$ connects qubits $q_i$ and $q_j$ that have two-qubit gates in between. 
To capture temporal information, we assign time-dependent weight to edges, where the weight $w_ij$ for the edge between $q_i$ and $q_j$ are defined by:
\begin{equation}
w_{ij} = \sum_{s=1}^{S} \alpha_s \cdot f_{ij}^{(s)}
\label{eq:weight}
\end{equation}
where S is the total number of segments, $ f_{ij}^{(s)} $ is the total number of two-qubit gates between qubits $q_i$ and $q_j$ within $s_th$ segment, and  $ \alpha_s $ is a time-dependent coefficient for segment s, with larger values assigned to segments in the near future. 
Specifically, we define $\alpha_s = \exp(-\lambda \cdot s) $, and $\lambda$ is a decay constant controller of the rate of the importance of future interactions diminishes. 

With this time-aware information, the interaction graph emphasizes qubits that will interact soon and ensure that they will be placed in near QPUs, and it also accounts for important interactions in later segments.
To adjust the hyperparameters will balance immediate and future communication needs. To assign qubits to each QPU, we perform clustering to partition $G$ into $P$ disjoint subsets $ \{V_1, V_2, \ldots, V_P\} $, and the goal is to minimize the total weight edge cut 
\begin{equation}
    C_{twe} = \sum_{\substack{e_{ij} \in E \\ q_i \in V_p, \ q_j \in V_{p{\prime}}, \ p \ne p{\prime}}} w_{ij}
\end{equation}
and the number of communities to be detected $k$ is given by the user, which denotes the QPU to be used. 
Although we assign logical qubits to QPUs by minimizing the total weighted edge cut, this initial assignment may violate the capacity constraints of the QPUs. 
These violations will be addressed by a refinement step where we adjust the assignment in a greedy fashion: if a QPU exceeds its capacity, we move some qubits to adjacent QPUs with available capacity, prioritizing qubits that have weaker connections within their current QPU. 
This refinement step ensures that capacity constraints are always met with minimally increasing the communication overhead.

\subsection{Simulated Annealing}

Our choice of simulated annealing is motivated by the inherent characteristics of distributed quantum circuit optimization. The main goal is to balance various factors that affect the circuit's performance, creating a complex optimization landscape where the solution space is primarily determined by qubit assignment decisions. Simulated annealing is particularly well-suited for such problems, as it can effectively search this complex space. By allowing qubit movement between segments, we further improve performance because qubits can be reassigned to optimize interactions in different parts of the circuit. Within each segment, we map qubits in a way that enhances the overall performance of the distributed circuit. The performance of a distributed quantum circuit can be affected by several factors, including local overhead (e.g., intra-QPU gate execution and swaps), remote overhead (e.g., inter-QPU communication and qubit movements between QPUs), and the cost of qubit movements between segments.

We quantify the performance using a metric that involves these factors, and we employ a simulated annealing algorithm to refine the qubit-to-QPU assignments by minimizing the following objective function:

\begin{equation}
E = \gamma_1 E_{\text{inter}} + \gamma_2 E_{\text{local}} + \gamma_3 E_{\text{move}},
\label{eq:objective_function}
\end{equation}

where $E_{\text{inter}}$ represents the inter-QPU communication cost, $E_{\text{local}}$ denotes the intra-QPU operation cost, and $E_{\text{move}}$ accounts for the cost of qubit movements between QPUs.

The term  $E_{\text{inter}}$  is the cost associated with inter-QPU interactions and penalizes assignments where frequently interacting qubits are placed on different QPUs that lead to higher communication overhead. The intra-QPU operation cost  $E_{\text{local}}$ accounts for the overhead within each QPU defined by 
\begin{equation}
E_{\text{local}} = \sum_{p=1}^{P} \sum_{\substack{(q_i, q_j) \in G_p}} w_{ij} \cdot c_{\text{local}}^{(p)}(q_i, q_j),
\label{eq:local}
\end{equation}
where P is the total number of QPUs used in this quantum circuit,  $G_p$ is the set of two-qubit gates between qubits assigned to QPU  p, and $c_{\text{local}}^{(p)}(q_i, q_j)$  is the cost of performing a gate between  $q_i$  and  $q_j $ on QPU  p, considering the swap overhead by limited topology. $E_move$ defined by 
\begin{equation}
E_{\text{move}} = \sum_{q_i \in Q_{\text{move}}} c_{\text{move}}(q_i),
\label{eq:movement}
\end{equation}
represents qubit movement between segments, and  $c_{\text{move}}(q_i)$ is the cost of moving qubit  $q_i$ between QPUs, which may depend on the distance between QPUs and the quality of quantum links. $ \gamma_1, \gamma_2, \gamma_3$ are coefficients that balance the contribution of each term based on their importance. In our model, we assume the movement of qubits is done by quantum teleportation~\cite{pirandola2015advances} and inter-QPU operation is performed by cat-entangler and cat-disentangler~\cite{yimsiriwattana2004generalized}. 

These coefficients will be adjusted by many factors. 
For example, if the hardware has faster, higher-quality quantum interconnects, the performance degradation due to remote overhead will not be severe. 
Then, we can assign a smaller value to  $\gamma_1$ to reduce the emphasis on minimizing inter-QPU communication costs  $E_{\text{inter}}$. 
On the other hand, if local operations on certain QPUs are costly due to limited connectivity or higher error rates, we need to increase $ \gamma_2 $ to penalize intra-QPU overheads  $E_{\text{local}}$  more heavily. 
By carefully tuning these coefficients, we tailor the optimization to the specific characteristics of the DQC configuration.
This flexibility ensures that the annealing process effectively balances the trade-offs between local and remote overheads according to their actual impact on performance.

The simulated annealing algorithm iteratively searches for a qubit-to-QPU assignment that minimizes the objective function  $E$. 
Starting with an initial assignment, at each iteration, the algorithm processes a new assignment by making small changes (swapping multiple qubits) or moving a qubit to a different QPU with capacity constraints met. 
For each iteration, the change in the objective function $  \Delta E = E_{\text{new}} - E_{\text{current}} $ is calculated, and the new assignment is accepted with a probability based on the following criterion:
\begin{equation}
P = \begin{cases}
1, & \text{if } \Delta E \leq 0, \\
\exp\left( -\frac{\Delta E}{T_k} \right), & \text{if } \Delta E > 0,
\end{cases}
\end{equation}
\label{eq:cri}
where $T_k$ is the current temperature at iteration k.
The temperature $T_k$ is updated according to the cooling schedule. 
More specifically, we use the cooling scheduling with: $ T_{k+1} = T_0 / (1 + \alpha k)$, where $T_0$ is the initial temperature and $\alpha$ is the cooling rate. 

Our simulated annealing method takes advantage of the circuit patterns identified during segmentation to guide the optimization process. 
A common challenge with simulated annealing approaches is their sensitivity to the quality of initial solutions - poor starting points can lead to suboptimal results and slower convergence. 
We address this limitation by initializing the annealing process with high-quality placements derived from the result of the previous segment which provides a perfect initial solution for the next segment. 
\section{Evaluation}
\label{evaluation}




\subsection{Experimental Setup}
The experimental evaluation was conducted on a dual-socket Intel Xeon Silver 4314 system running Ubuntu Linux 22.04.1 LTS. Each socket contains a 16-core CPU supporting hyper-threading, providing 32 physical cores and 64 logical CPUs in total. The server has 125 GB of RAM and 8 GB of swap memory. All benchmark circuits were generated using Qiskit~\cite{qiskit2024} version 0.45.0 and collected from~\cite{li2023qasmbench}. We evaluated three types of quantum circuits with varying qubit counts: Quantum Fourier Transform (QFT), Quantum Approximate Optimization Algorithm (QAOA), and quantum arithmetic circuits. To simulate a heterogeneous QPU cluster, we created a random topology using multiple fake backends (FakeAuckland, FakeParis, FakeMelbourne, and FakeBoeblingen). Each simulated QPU's computational qubits follow the coupling map defined by its fake backend. Additionally, we implemented a quantum switch that controls entanglement resource distribution. The quantum switch of a certain QPU maintains full connectivity with its computational qubits.

\subsection{Main Result}
In this section, we demonstrate the effectiveness and functionality of our method. The results are presented in Table~\ref{tab:objective_reduction}. We set the method of finding a feasible assignment of qubits with random placement as the baseline and compare the objective value reduction achieved by different layers of our approach: L1: Utilizes Time-Aware clustering with circuit segmentation but without annealing;L2: Employs Time-Aware clustering and annealing but with random circuit segmentation, without detecting patterns; L3: Represents our complete method, which includes segmenting circuits by detecting patterns, along with Time-Aware clustering and annealing. 
The primary observation is that, for most of the circuits we tested, the objective value reduction improves as we incorporate more layers of our method, which validates the effectiveness of our overall design. 
Another notable observation is that the relative improvement ratios of different layers vary across circuits. This variation can be attributed to the inherent characteristics of the circuits. QFT circuits exhibit distinct circuit patterns and qubit locality, and this characteristics make pattern detection more beneficial. In contrast, QAOA circuits have a repeated pattern across the complete set of qubits, so pattern-based segmentation offers less additional advantage. Therefore, L2 segmentation yields varying levels of improvement, offering more in QFT circuits and less in QAOA circuits. In most cases, our annealing method (used in L2 and L3) further reduces the overhead by optimizing the qubit mapping to balance inter-QPU and intra-QPU communications effectively. Overall, the incremental improvements from L1 to L3 across different circuits validate the effectiveness of our layered approach in enhancing mapping efficiency.


\begin{table}[htbp]
    \caption{Objective Value Reduction by Layers.}
    \centering
    \setlength{\tabcolsep}{4pt} 
    \renewcommand{\arraystretch}{1.1} 

    \begin{tabular}{cccc}
        \toprule
        \textbf{Program} & 
        \textbf{L1 Red.} & 
        \textbf{L2 Red.} & 
        \textbf{L3 Red.} \\
        \midrule
        qft\_100   & 8.47\% & 27.84\% & 34.73\% \\
        qft\_160   & 7.05\% & 7.69\% & 14.24\% \\
        qft\_320   & 7.79\% & 10.87\% & 12.42\% \\
        qaoa\_100  & 94.19\% & 92.58\% & 94.33\% \\
        multiplier\_75  & 59.71\% & 57.18\% & 66.44\% \\
        adder\_n118 & 84.78\% & 85.04\% & 88.40\% \\
        \bottomrule
    \end{tabular}
    \label{tab:objective_reduction}
\end{table}
\vspace{-2ex}
\subsection{Evaluation with Different Inter-QPU and Intra-QPU Overhead Ratios}

\begin{figure}

    \centering
    \includegraphics[width=0.9\linewidth]{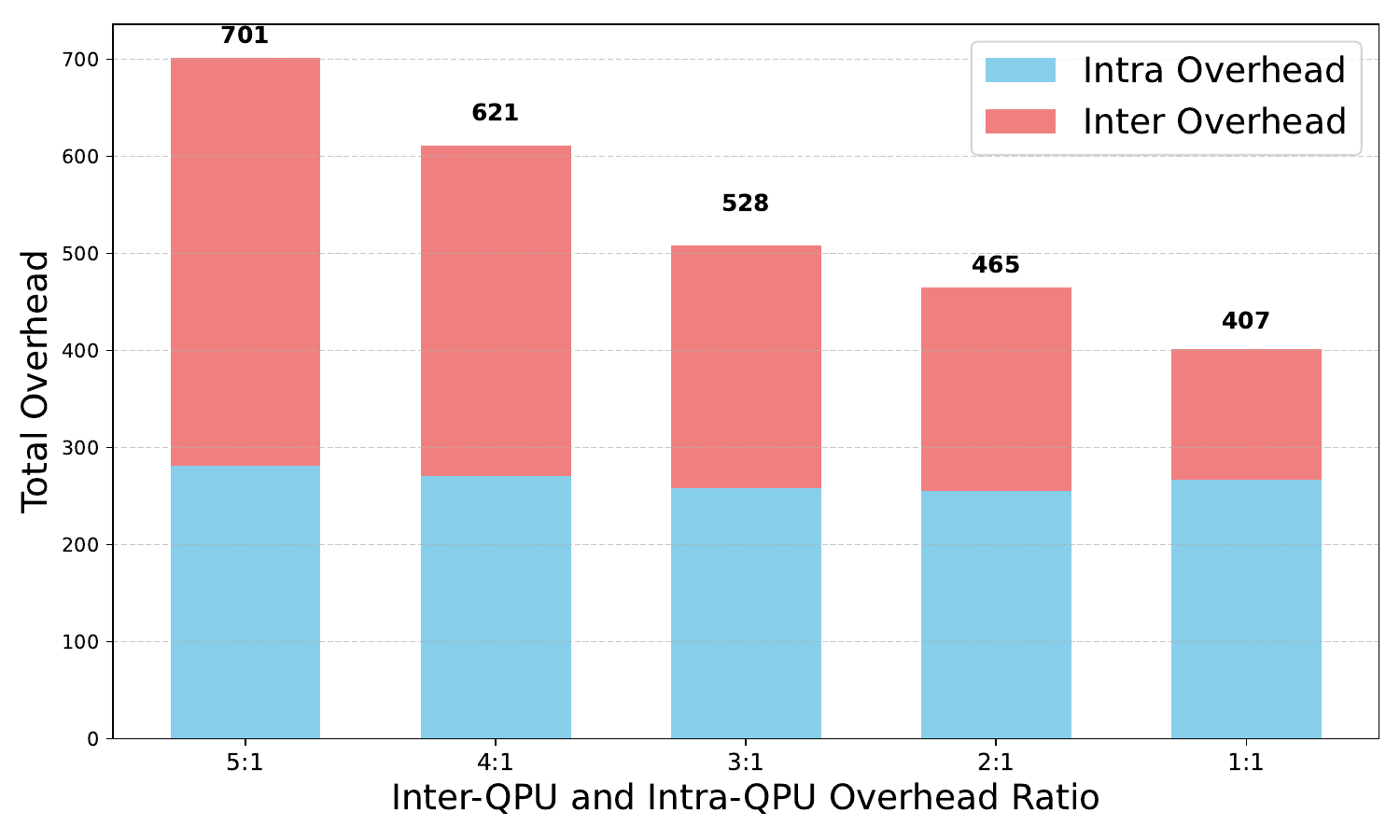}
    \caption{Total Overhead with Different Inter-QPU and Intra-QPU overhead ratio}
    \label{fig:interconnect}

\end{figure}

In this subsection, we aim to test the effectiveness of our method under different ratios of Inter-QPU to Intra-QPU overheads. We vary the ratio from 5:1 to 1:1, which aligns the development of quantum interconnects where advancements lead to faster inter-QPU communication. In this evaluation, we consider both Inter-QPU operations and movement as Inter-QPU overhead since they both rely on quantum inter-connects. We perform our evaluation on the \texttt{adder\_n118} circuit, and the results are shown in Fig.~\ref{fig:interconnect}. From the results, we observe that as quantum interconnects become faster (i.e., the Inter-QPU overhead decreases relative to the Intra-QPU overhead), the total energy consumption decreases. Specifically, the intra-QPU overhead remains relatively unchanged across different ratios, while the portion of inter-QPU overhead decreases with faster interconnects. This trend indicates that as inter-QPU communication becomes more efficient and its overhead becomes comparable to intra-QPU overhead, the inter-QPU overhead contributes less to the total overhead. However, this observation aligns with our motivation that in future distributed quantum computing (DQC) systems, solely minimizing the inter-QPU overhead is not enough. Instead, we need to take various factors that affect the performance of DQC. Considering that quantum algorithms are continuously evolving, different quantum circuits will exhibit distinct characteristics, which may further influence this balance. Our method's ability to adapt to varying overhead ratios demonstrates its effectiveness in optimizing performance across different DQC scenarios.

\section{Conclusion}
\label{conclusion}
In this work, we design a new compilation framework for distributed quantum computing. Our framework comprises three major steps: circuit pattern detection, time-aware clustering, and simulated annealing, each addressing specific challenges in the compilation process. The primary contribution of our work is the consideration of the patterns and characteristics inherent in quantum circuits, which are critical to the performance of DQC. As quantum technology advances, we will inevitably enter a heterogeneous QPU environment where multiple factors must be balanced rather than solely minimizing remote communication overhead. We believe these challenges are imminent, and our work aims to address them and fill the gap. Our experimental results demonstrate the effectiveness(88.40\% reduction) of our methods in handling the complexities of heterogeneous distributed quantum systems.




\clearpage

\bibliographystyle{IEEEtran}  
\bibliography{ref}         

\end{document}